\documentclass[apj]{emulateapj}

\shorttitle{Mass-dependent Color Evolution of Galaxies}
\shortauthors{Kajisawa and Yamada}

\begin{document}


\title{Mass-dependent Color Evolution of Field Galaxies Back to
$z\sim3$ Over the Wide
Range of Stellar Mass}

\author{M. Kajisawa 
and T. Yamada
}
\affil{National Astronomical Observatory of Japan,\\
2-21-1, Osawa, Mitaka, Tokyo 181-8588, Japan}
\email{kajisawa@optik.mtk.nao.ac.jp}

\begin{abstract}
We use deep multi-band optical and near-infrared data for four general
fields, GOODS-South, HDF North/South, and IRAC UDF in GOODS-North to
investigate the evolution of the observed rest-frame $U-V$ color of
field galaxies as a function of the stellar mass evaluated by fitting
the galaxy spectral models to the observed broad-band SEDs. In these
four fields, the $U-V$ color distributions of the galaxies at each mass
and redshift interval are very similar. At $0.3 < z < 2.7$, we found
that more massive galaxies always tend to have a redder $U-V$
color. High- and low-mass galaxies exhibit quite different color
evolutions. As seen in our previous study in HDF-N, the color
distribution of low-mass (M$_{stellar} \lesssim 3 \times 10^{9}$M$_{\odot}$)
galaxies becomes 
significantly bluer with an increase in the redshift. This evolution
of the average color can be explained by a constant star formation
rate model with $z_{\rm form} \sim 4$. On the other hand, the average color of
high-mass galaxies (M$_{stellar} \gtrsim 3 \times 10^{10}$ M$_{\odot}$)
evolves more strongly at 
a high redshift. Such mass-dependent color distribution and its
evolution indicate that galaxies with a larger stellar mass appear to
have shorter star-formation timescales, and on an average, they form
the larger fraction of their stars in the earlier epoch.
\end{abstract}




\section{Introduction}
 Stellar mass is one of the most important quantities of galaxies that
 facilitate the understanding of their formation and evolution. The
 evolution of the number density of galaxies as a function of the
 stellar mass conveys the overall history of star formation and the
 mass assembly process in the universe. Many authors have investigated
 such evolution of the stellar mass distribution function back to 
$z\gtrsim1$ (e.g., \citealp{dic03}, 
\citealp{bel03}, \citealp{fon04}, \citealp{dro04}, \citealp{dro05}).


 On the other hand, the study of properties such as the spectral
 energy distribution (SED), colors, and line fluxes of galaxies as a
 function of the stellar mass at various epochs helps us to understand
 star formation histories in additional detailed, such as, in what
 manner the star formation occurs in low- and high-mass galaxies and
 how it contributes to the growth of their stellar mass. In the case
 of local galaxies detected in the Sloan Digital Sky Survey, 
\cite{hea04} and \cite{jim05} argued that most of the stars
 in the present-day massive galaxies have been formed at a high
 redshift, while star formation histories of low-mass galaxies have
 relatively longer timescales. In more direct observations of
 high-redshift galaxies, \cite{cow96} originally found that the
 mass of the galaxies undergoing rapid star formation has been
 declining smoothly with a decrease in the redshift. Recently,
 \cite{jun05} also investigated the stellar mass dependence of star
 formation rates of galaxies at $0.8 < z < 2.0$ using the Gemini Deep
 Deep Survey data. They found the trend to be similar to that
 predicted in the studies of local galaxies \citep{hea04}. 
Similar results are obtained in the studies on the
 evolution of specific star formation rates back to $z \sim $1-1.5
 (\citealp{bri00}, \citealp{bau05}, \citealp{feu05a}). 
Very recently, \cite{feu05b} extended the
 analysis to $z \sim 5$, and found that a similar trend continues even at a
 higher redshift.

The samples in these previous studies, however, are too shallow to
reach a sufficiently low mass range below $\sim 5 \times 10^{9}$
M$_{\odot}$, a
characteristic ``transition'' mass (Kajisawa and Yamada 2005, and
Section 4 below), or are biased toward objects with typically low M/L
ratios, such as more actively star-forming bluer populations 
 due to the lack of very deep near-infrared
(NIR) data. The evolution of low-mass galaxies is, however, important
to understand the overall picture of galaxy evolution not only because
they may be the building blocks of more massive galaxies but also
because they are more sensitive probes to understand the thermal
history of the universe.

 In Kajisawa and Yamada (2005, hereafter referred to as KY05), we
 investigated the evolution of the rest-frame color and morphology
 distribution of the galaxies in HDF-N as a function of the stellar
 mass back to $z \sim 3$ over a wide range of stellar masses down to
 $10^{9}$ M$_{\odot}$. We found that the behavior of the
 rest-frame $U-V$ color distribution and its dependence on the stellar
 mass change at around $\sim 5 \times 10^{9}$ M$_{\odot}$. 
In addition, we discovered that
 the sequence of the color distribution of low-mass galaxies
 systematically becomes bluer with the redshift, while the colors of
 high-mass galaxies do not show a significant redshift
 evolution. However, the area of the HDF-N data used in our analysis
 was only $\sim$ 3.5 arcmin$^{2}$, and it was inferred that the effect of
 field-to-field variation was large.

\begin{figure*}
\vspace{10mm}
\hspace{15mm}
\includegraphics[angle=90, scale=1.1]{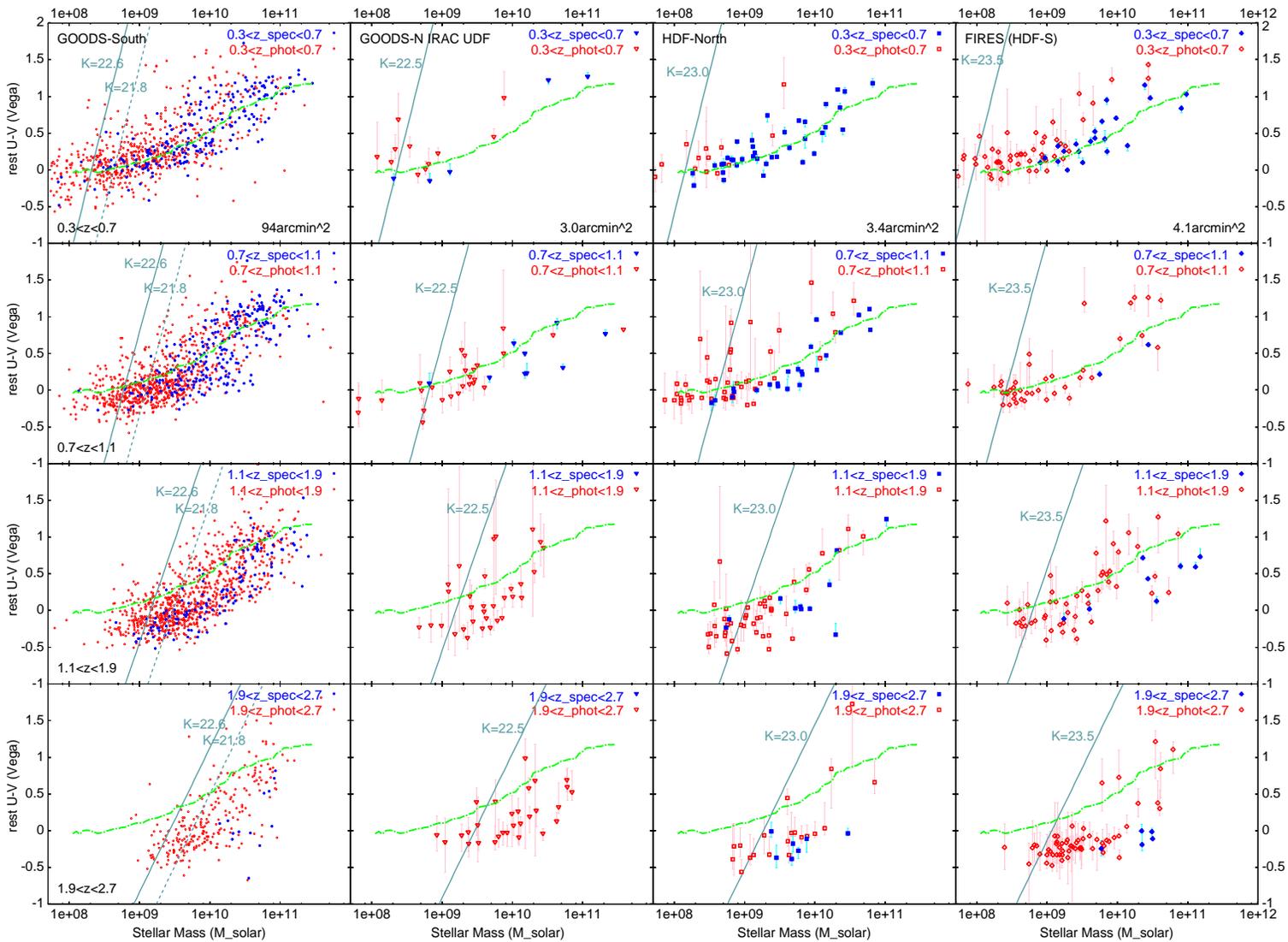}
\caption{Rest-frame $U-V$ color distribution as a function of stellar mass
of $K$-selected galaxies.
Each row depicts different redshift bin, and each column depicts
different fields. Blue symbols indicate objects with spectroscopic redshift,
while red ones indicate the phot-z sample.
Error bars show the range of 90\% confidence level (the redshift
uncertainty is considered for the phot-z sample).
Solid lines (and dashed lines for the most shallow ISAAC field in 
GOODS-S) show the detection limit that corresponds to the 90\%
completeness limit for point sources (see KY05 for the method of
estimating the lines). The dotted-dash lines show the weighted average for
the GOODS-S sample at $0.3<z<0.7$ (top-left panel).
\label{MsUVall}}
\end{figure*}

 Therefore, in this paper, we have extended the study area of KY05 to
 include some other fields, GOODS-S \citep{gia04}, 
a part of the IRAC Ultra Deep Field in GOODS-N \citep{dic03b}, and
 FIRES (HDF-South, \citealp{lab03}). 
Using the deep optical and NIR data for these fields, we
 investigate the rest-frame $U-V$ color evolution of galaxies at $0.3 < z
 < 2.7$ as a function of the stellar mass. These data allow us to
 extend our previous analysis to a larger sample in order to study the
 evolution of galaxies with a high statistical accuracy. We used the
 Vega magnitude system throughout the paper.

\begin{figure*}[t]
\epsscale{1.1}
\hspace{-20mm}
\plotone{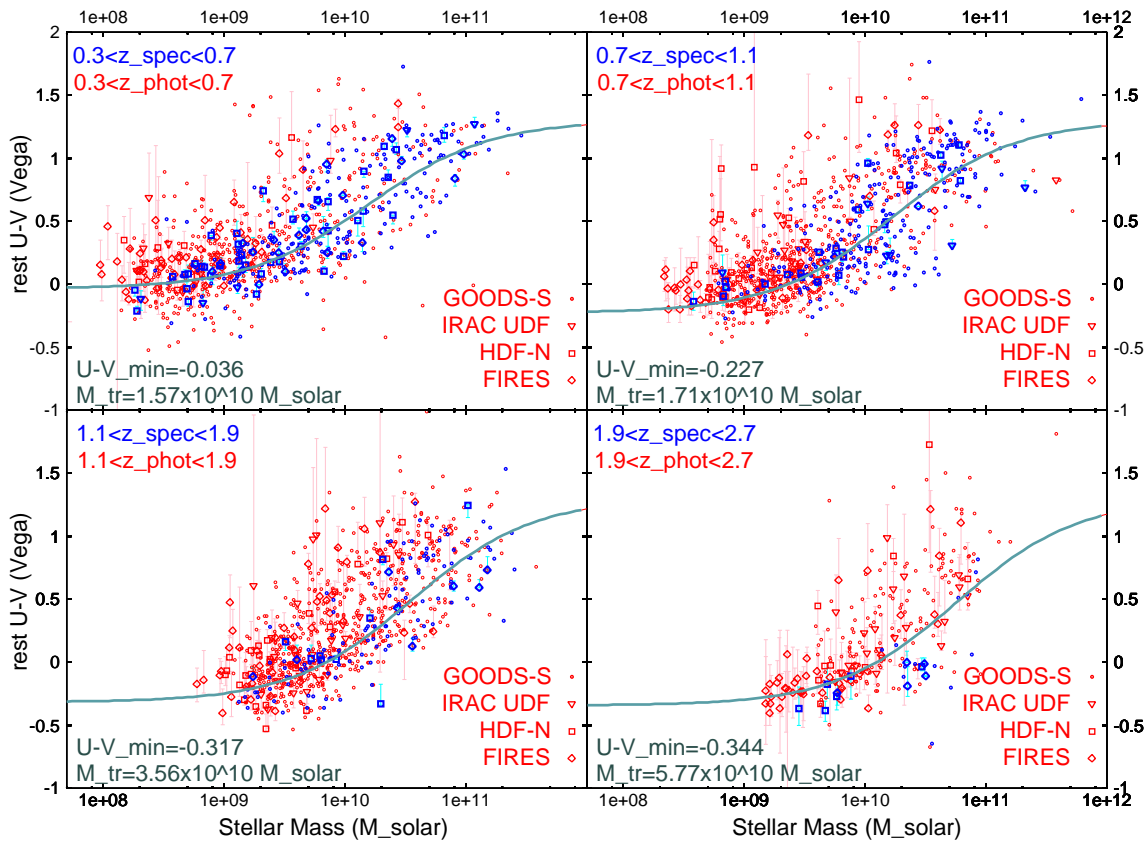}
\caption{Rest-frame $U-V$ color distribution as a function of 
the stellar mass
for the sum of all the four fields in each redshift bin. 
Data from different fields are plotted by different symbols.
Blue and red symbols represent
the spec-z and phot-z samples, respectively. Solid lines show the
fitted ``tanh'' function in each redshift bin. The fitting parameters,
UV$_{MIN}$ and M$_{tr}$ (see text) are also shown.
\label{MsUVfit}}
\end{figure*}

\section{Data \& Source Detection}

We used multi-band optical and NIR data in the four fields, GOODS-S,
FIRES, HDF-N, and a part of IRAC UDF in GOODS-N. For HDF-N, we used
the same data as those used in KY05 (HST/WFPC2, HST/NICMOS,
Subaru/CISCO, UBVIJHK set). For GOODS-S, we used the publicly
available VLT/ISAAC ver1.5 data (JHK-bands) and HST/ACS ver1.0 data
(bviz, Giavalisco et al. 2004). For FIRES, we used the publicly
available HST/WFPC2 $+$ VLT/ISAAC data (UBVIJHK, Labbe et al. 2003).


 For IRAC UDF in GOODS-N, we observed a field in the $J$ and $K$ bands
 using the Subaru Telescope equipped with CISCO (Motohara et al. 2002)
 on March 10, 2003. The field of view (FOV) used in the analysis was
 1.74' $\times$ 1.74', and the total exposure time in the $K$ and $J$ bands was
 14780 and 3880 s, respectively. The data were reduced using the IRAF
 software package, as described in KY05. The publicly available GOODS
 HST/ACS ver1.0 data (bviz) were also used in the field.

 For all the fields, we independently performed source detection in
 the $K$-band images using the SExtractor image analysis package
 \citep{ber96}.  We adopted MAG\_AUTO from SExtractor as the total
 magnitude of the detected objects. The SED (colors) was measured in
 the optimal-S/N aperture size for each galaxy (see KY05 for
 details). We excluded regions with lower sensitivity, such as the
 edge region of each ISAAC FOV for GOODS-S, from our analysis. The
 areas in the analyses were 94 arcmin$^{2}$ for GOODS-S, 3.0 arcmin$^{2}$ for
 IRAC UDF, 3.4 arcmin$^{2}$ for HDF-N, and 4.1 arcmin$^{2}$ for FIRES. We
 performed simulations using the IRAF/ARTDATA package to quantify the
 depth of each field. The 90\% completeness limit for the point source
 is $K \sim $21.8-22.6 ($\sim$21.4 for one FOV) for GOODS-S, $K \sim
 22.5$  for IRAC UDF, $K \sim 23.0$ for HDF-N, and $K \sim 23.5$ for FIRES.

\section{Analysis}

Using these optical and NIR photometric data, we measured the
redshift, the stellar mass, and the rest-frame $U-V$ color of each
galaxy in the same manner as that in KY05. When possible, we used the
spectroscopic redshifts from literature (\citealp{szo04}, \citealp{lef04}, 
\citealp{van05}, \citealp{wir04}, \citealp{coh00},
\citealp{lab03}). For objects 
without spectroscopic identification, we estimated the photometric
redshifts using the publicly available ``hyperz'' code (Bolzonella et
al. 2000) in the same manner as that in KY05; however, for FIRES, we
used the publicly available phot-z catalogue \citep{lab03}.

 After the redshifts were determined, we performed a detailed SED
 fitting with the GALAXEV synthetic library \citep{bru03} in
 order to measure the stellar mass and the rest-frame color of the
 galaxies.  We assumed IMF of \cite{cha03},
  exponentially decaying SFR, and the
 extinction law of \cite{cal00}. The free parameters were
 the age, star formation timescale, color excess, and metallicity. The
 fitting results were used to estimate the M/L ratio (and
 subsequently, the stellar mass) and measure the K-correction values
 for the rest-frame colors. The parameter range of dust extinction is
 $0 \le$ E($B - V$) $\le 0.62$, and the median of the best-fitted
 values is 0.13 
 (about 75\% of the samples had E($B - V$) $< 0.3$). Additional details of
 the procedures and the discussion on the uncertainty of the mass and
 color evaluation are described in KY05. It should be noted that
 although each parameter such as dust extinction, stellar age, or star
 formation timescale is not constrained very tightly, the stellar mass
 can be estimated with a relatively low uncertainty. This is because
 the effects of these parameters tend to cancel out in the estimation
 of the stellar mass, as discussed in KY05 (see also \citealp{pap01}
  for a detailed discussion). In addition, note that the
 rest-frame $U-V$ color is measured from the observed photometric data
 without extrapolation. Therefore, the stellar mass and the $U-V$ color
 used in the current analysis are robust quantities.

\section{Results \& Discussion}

 In Figure \ref{MsUVall}, for each of the four fields, we plotted
 the rest-frame 
 $U-V$ color of the K-selected galaxies in each redshift bin as a
 function of the stellar mass. First, we can see from this figure that
 the color distributions of the galaxies in the four fields are very
 similar. For reference, we estimated the weighted mean colors with a
 bin of $\pm$0.15 dex of the GOODS-S sample at $0.3 < z < 0.7$ (in the
 top-left panel), and plotted them in all the panels by dotted-dash
 lines. The $U-V$ color is always very blue for M$_{stellar} \lesssim
 10^{9}$ M$_{\odot}$ 
 and becomes redder as the stellar mass increases toward $\sim
 10^{11}$ M$_{\odot}$. 
Most of the galaxies with M$_{stellar} \gtrsim 10^{11}$ M$_{\odot}$
have the reddest color. 

In fact, it is evident from Figure \ref{MsUVall} that over the entire
redshift range at $0.3 < z < 2.7$, the more massive galaxies always have
the redder $U-V$ color. For example, while most of the galaxies with
M$_{stellar} \lesssim 3 \times 10^{9}$ M$_{\odot}$ have $U-V < 0.3$ at
any redshift, many objects
with M$_{stellar} \gtrsim 3 \times 10^{10}$ M$_{\odot}$ have $U-V \sim 
1$. This trend is consistent 
with the results found in \cite{feu05b} and other previous
studies. If we observe the redshift evolution in Figure \ref{MsUVall} more
closely, we notice the gradual bluing of the $U-V$ color along the
redshift within the mass range of $10^{9}$ M$_{\odot} \lesssim$
M$_{stellar} \lesssim 10^{10}$ M$_{\odot}$. While
this systematic evolution of the rest-frame color distribution of
low-mass galaxies has already been reported in KY05 for HDF-N, it is
now confirmed by the larger sample that includes three other different
fields.

 In Figure \ref{MsUVfit}, we try to quantify the observed
 mass-dependent color 
 evolution by using an analytical fitting function. The data for all
 the four fields are combined. We excluded the lowest mass galaxies in
 each field from the sample so that the color bias near the detection
 limit (solid lines in Figure \ref{MsUVall}) does not affect the fitting
 significantly; the limiting masses are set so that the $K$-band source
 detection completeness at $U-V \sim 0.8$ is 50\%. These limiting stellar
 masses are tabulated in Table \ref{cutMs}.
Considering the shape of the mass-weighted average color profile in
 GOODS-S (the top-left panel in Figure 1), we chose a hyperbolic
 tangent as the fitting function. This was in keeping with \cite{bal04}
  who used a similar form to quantify the bimodality in the
 color-magnitude distribution of the SDSS sample. The fitting function
 is written as follows:
\begin{table}
\caption{Lower limit of stellar mass in each field
(M$_{\odot}$)
\label{cutMs}}
\begin{tabular}{ccccc}
\tableline\tableline
Redshift & GOODS-South & IRAC UDF & HDF-N & FIRES\\
\tableline
0.3-0.7 &1.4-3.3$\times 10^{8}$ &1.8$\times
10^{8}$ &1.1$\times 10^{8}$ &6.6$\times
10^{7}$\\
0.7-1.1 &4.2-10$\times 10^{8}$ &5.5$\times
10^{8}$ &3.5$\times 10^{8}$ &2.0$\times 10^{8}$\\
1.1-1.9 &1.2-2.9$\times 10^{9}$ &1.6$\times
10^{9}$ &1.0$\times 10^{9}$ &5.8$\times 10^{8}$\\
1.9-2.7 &3.1-7.6$\times 10^{9}$ &4.1$\times
10^{9}$ &2.6$\times 10^{9}$ &1.5$\times
10^{9}$ \\
\tableline
\end{tabular}
\end{table}
\\
\vspace{1mm}
$U-V$$({\rm
M}_{stellar})$=$
\left(\frac{UV_{MIN}+UV_{MAX}}{2}\right)+\\\left(\frac{UV_{MAX}-UV_{MIN}}{2
}\right)$$\times\tanh{(R({\rm
M}_{stellar}-{\rm M}_{tr}))}$
\vspace{1mm}
\\
where $UV_{MIN}$ and $UV_{MAX}$ are the asymptotic minimum and maximum
values of 
the rest-frame $U-V$ color at the low- and high-mass ends.
M$_{tr}$ and $R$ represent the mass and mass range at which the color
transition occurs. For simplicity, we fixed $R$ and $UV_{MAX}$, which seem to
change little with the redshift in our sample, to the values fitted at
$0.3 < z < 0.7$ ($R = 1.0$, $UV_{MAX} = 1.3$). Next, we fitted the color
distribution in each redshift bin by changing the values of M$_{tr}$ and
$UV_{MIN}$. The result is shown by the solid lines in Figure
\ref{MsUVfit}. The
best-fitted values of the parameters are M$_{tr} =
1.57^{+0.39}_{-0.23} \times 10^{10}$
M$_{\odot}$ and
$UV_{MIN} = -0.036 \pm 0.055$ at $0.3 < z < 0.7$, M$_{tr} =
1.71^{+0.28}_{-0.24}
\times 10^{10}$ M$_{\odot}$ and
$UV_{MIN} = -0.277 \pm 0.056$ at $0.7 < z < 1.1$, M$_{tr} =
3.56^{+1.09}_{-0.84}
\times 10^{10}$ M$_{\odot}$ and
$UV_{MIN} = -0.317 \pm 0.128$ at $1.1 < z < 1.9$, M$_{tr} =
5.77^{+1.79}_{-1.37}
\times 10^{10}$ M$_{\odot}$ and
$UV_{MIN} = -0.344 \pm 0.084$ at $1.9 < z < 2.7$.
These results again show that the $U-V$ color at the low-mass end
gradually evolves blueward with the redshift. In addition, the
transition mass also seems to increase along the redshift, although
the uncertainty is relatively large. It is noteworthy that \cite{bun06}
 also recently discovered similar transition mass increments
with the redshift at $0.4 < z < 1.4$ in the study using the DEEP2 data.

\begin{figure*}
\epsscale{0.9}
\hspace{-30mm}
\plotone{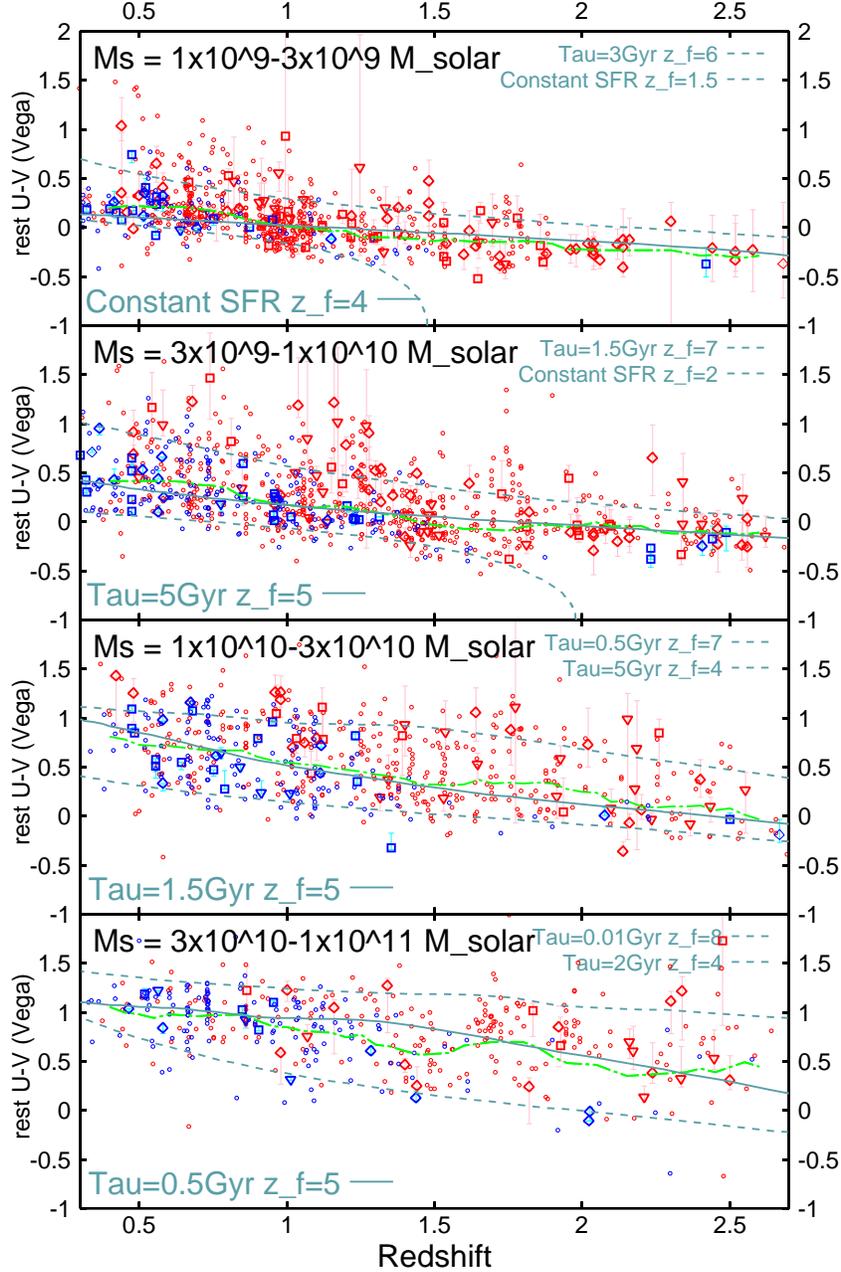}
\caption{rest-frame $U-V$ color vs redshift for galaxies within each 
stellar mass range. The plotted sample and the meaning of the symbols
are 
similar to those in Figure \ref{MsUVfit}.
Dotted-dash lines and solid lines show the weighted
average and the best-fit GALAXEV models with exponentially
decaying SFR. 
Dashed lines show the upper and lower envelope of the
color distribution.
\label{UVz}}
\end{figure*}

Figure \ref{UVz} shows the evolution of the rest-frame $U-V$ color at each
stellar mass range for the combined data. We plotted the weighted
average colors binned over the redshift range of $\pm$0.15 (dotted-dash
lines). In the top panel, it is more clearly evident that the color
distribution of the galaxies with M$_{stellar} < 3 \times 10^{9}$
M$_{\odot}$ becomes 
gradually bluer with an increase in the redshift. In KY05, we
discussed that the star formation activity in these low-mass galaxies
occurs rather continuously since they exhibit a very blue color at
{\it any}
redshift. A bluer color even at a higher redshift also suggests
younger average ages of the galaxies at $z =$2-3. In the lower panels,
we see that the colors of galaxies with a larger stellar mass also
become bluer at a high redshift. There seems to be a trend where the
more massive populations show a stronger color evolution at a high
redshift, which in fact corresponds to the evolution of the transition
mass found in Figure \ref{MsUVfit}.

 We try to fit such color evolutions using GALAXEV population
 synthesis models with simple star formation histories. We simply
 consider an exponentially decaying SFR and assume solar metallicity
 and no dust extinction. We estimate the $U-V$ color for the model grid
 with the star formation timescale $\tau=$ 0.01, 0.05, 0.1, 0.5, 1, 1.5,
 2, 3, 5, 15, and $\infty$ (constant SFR) Gyr and the formation
 redshift $z_{f} =$ 3, 4, 5, 6, 7, and 8. 
Eventually, we found that $\tau$ mainly affects
 the overall slope of the color evolution, and $z_{f}$ strongly affects the
 color at a high redshift (e.g., $z > 2$). The resultant best-fit models and
 their $\tau$ and $z_{f}$ values are shown in Figure \ref{UVz}. 
It is found from the
 results that the more massive populations appear to have a shorter
 star formation timescale. In other words, they form the larger
 fraction of their stellar contents at a high redshift. Such a trend
 is very similar to the concept of mass-dependent star formation
 histories argued by \cite{hea04}. Further, the trend is also
 consistent with the studies on mass-dependent evolutions of specific
 star formation rates at a high redshift (\citealp{feu05b},
 \citealp{jun05}). 
However, in this study, the trend is depicted by more
 directly observable quantities.

 In these analyses, it is important to check the effect of the color
 bias near the detection limit. The bias against a red high-M/L
 population is expected to be severe, especially for low-mass galaxies
 at a high redshift since the $K$-band filter samples relatively short
 rest-frame wavelengths. We finally discovered, however, that our deep
 $K$-selected samples do not suffer considerably from the bias. It is
 true that our 50\% completeness limit at $U-V = 0.8$ could neglect some
 fractions of relatively red low-mass galaxies in the highest redshift
 bin. However, the color distribution tends to concentrate at
 considerably bluer values ($U-V \lesssim 0$), and there is few galaxies
 that are redder than $U-V = 0.3$, where the completeness is greater
 than 90\%. This can be demonstrated by the lowest mass objects near
 the detection limit ($K \sim 23.5$) in the highest redshift bin of the
 deepest FIRES data.
For example, out of 20 galaxies with the stellar mass of $1.5 \times
 10^{9}$ M$_{\odot}
  <$ M$_{stellar} < 3.0 \times 10^{9}$ M$_{\odot}$ in the $1.9 < z <
 2.7$ bin of FIRES, 19
 galaxies have $U-V < 0$, and one galaxy has $0 < U-V < 0.1$; further, no
 galaxy has $0.1 < U-V < 0.3$.

 Indeed, a sample selection with very deep NIR data is essential not
 only for evaluating the stellar mass of galaxies but also for
 generating relatively unbiased samples to study the color
 distribution of these low-mass galaxies at a high redshift. For
 example, $I$-band selections such as those used for the FORS Deep Field
 data in \cite{feu05b} could introduce a severe color
 bias. The detection limit of $I(AB) = 26.4$ in \cite{feu05b} can
 only sample galaxies with $U-V < -0.1$ for M$_{stellar} = 2 \times
 10^{9}$  M$_{\odot}$ at $z
 > 2$. Such a selection does not reveal whether red galaxies exist.

In Figure \ref{UVz}, we found that the stronger color evolution of massive
galaxies at a high redshift was explained by the exponentially
decaying SFR models with $z_{f} \sim $4-5, for which the star
formation timescale decreases with mass. On the other hand, the
evolution of the average color of low-mass ($\lesssim 3 \times 10^{9}$
M$_{\odot}$) galaxies can 
actually be explained by the constant SFR model.
However, the best-fit formation redshift occurs at $z_{f} \sim 4$, which is
similar to those for high-mass galaxies. If this is true, the
differences in the star formation histories among galaxies with
different stellar masses are not mainly in the epochs when the star
formation began, but rather, in the star formation efficiencies and/or
mechanisms by which star formation activities are suppressed. The long
star-formation timescale for low-mass galaxies can be caused by the
self-regulation of star formation, for example, by a supernova
feedback \citep{dek05}. The mass dependence of the star
formation efficiencies and suppression mechanisms can be related with
processes such as the efficient exhaustion of gas reservoirs (e.g.,
\citealp{men05}) or shock heating of cold gas during a major merger
\citep{cox04}.
 
 It is noteworthy that some of the assumptions in our analysis may be
 oversimplifications. If we consider dust extinction, the estimated
 formation redshift shifts to smaller values. For example, if we
 assume the extinction of E($B - V$) $= 0.13$ (median-fitted value of our
 sample) for low-mass galaxies, the best-fit formation redshift of
 these galaxies shifts from $z_{f} = 4$ to $z_{f} = 3$. We also verified which
 variations of star formation histories are permissible if the scatter
 of the colors around the average values is considered. To illustrate
 this, we plotted the GALAXEV models that trace the upper and lower
 envelopes of the color distribution by dashed lines in each panel in
 Figure \ref{UVz}.

 Finally, we note that several other surveys have investigated the
 evolution of the stellar mass or luminosity function of field
 galaxies and found that the normalization decreases at $z \gtrsim 1$ (e.g.,
 \citealp{fon04}, \citealp{yam05}, \citealp{cap06}, \citealp{sar06}). Several
 studies have also suggested mass-dependent number density evolution
 (e.g., \citealp{gla04}, \citealp{dro04}, \citealp{dro05}). 
The next step is to
 consistently explain both the color (SED) evolution and the number
 density evolution as functions of the stellar mass.

\acknowledgments
We wish to thank the anonymous referee for invaluable comments.
This study is based on data collected at Subaru Telescope, which is operated by
the National Astronomical Observatory of Japan.
This study is also based on data collected at Very Large
Telescope at the ESO Paranal Observatory under Program ID:
LP168.A-0485. 
Data reduction/analysis was carried out on ``sb'' computer system
operated by the Astronomical Data Analysis Center (ADAC) and Subaru
Telescope of the National Astronomical Observatory of Japan.

\end{document}